\documentstyle[floats,twocolumn,prb,aps]{revtex}

\begin{document}

\title{Universal conductance
in quantum wires in the presence of Umklapp scattering}

\author{Satoshi Fujimoto$^{1,2}$ and Norio Kawakami$^3$}
\address{$^1$Department of Physics, Theoretical Physics, University of
Oxford, 1Keble Road, Oxford, OX1 3NP, U.K. \\ 
$^2$Department of Physics, Kyoto University, Kyoto 606, Japan \\
$^3$Department of Applied Physics,
Osaka University, Suita, Osaka 565, Japan}

\date{\today}

\maketitle

\begin{abstract}
The effects of Umklapp scattering on the zero-temperature conductance
in one-dimensional quantum wires are reexamined
 by taking into account both 
the screening of external potential and the non-uniform chemical
potential shift due to electron-electron interaction.
It is shown that in the case away from half-filling
the conductance is given by the universal value, 
$2e^2/h$, even in the presence of Umklapp scattering, 
owing to these renormalization effects of
external potential. The conclusion is in accordance with
the recent claim obtained for the system with 
non-interacting leads being attached to a quantum wire.
\end{abstract}

\section{Introduction}

The effects of electron-electron interaction on
the conductance in one-dimensional (1D) 
quantum wires have been extensively studied 
from both theoretical and experimental points of view.
The low-energy properties of 
1D interacting electron systems are described by the
Tomonaga-Luttinger (TL) liquids\cite{hal}. 
It has been known that the conductance at zero
temperature is
given by $2e^2K_{\rho}/h$, where $K_{\rho}$ is the TL liquid
parameter which controls the asymptotic behavior of correlation
functions\cite{ap,kane}.
However, according to the recent experiment, the observed
conductance is not $2e^2K_{\rho}/h$ but $2e^2/h$, which is 
expected for the conductance of
 1D non-interacting electron systems\cite{taru}.
In order to explain this discrepancy,
two possible scenarios have been proposed.
One is that non-interacting leads attached to
a quantum wire are essential to reproduce the observed
conductance $2e^2/h$\cite{stone,pome,schu}.
The other scenario is that 
if one takes into account the screening of external
potential due to electron-electron interaction, 
the multiplicative factor 
$K_{\rho}$ in the conductance may become unity\cite{kawa2,fink,shimi}.
Both of these scenarios give the nice explanations 
for the experimentally observed conductance, 
if only forward scattering of electron-electron interaction
exists.
However, in the presence of Umklapp or impurity scattering
which gives rise to momentum dissipation,
these scenarios lead to different 
results\cite{fink,maslov,furu,fk,naga,gra,toku,mori,mas,safi}.
In the case that non-interacting leads are attached,
the conductance is still given by the 
unrenormalized value, $2e^2/h$, even 
in the presence of Umklapp scattering
as far as the electron density is away from half-filling\cite{mori,mas}.
In contrast, if one takes into account only the screening
of external potential due to electron-electron interaction,
the Umklapp scattering gives the non-universal value
of conductance, $2e^2\gamma/h$, where $\gamma$ is a constant different 
from $K_{\rho}$\cite{fk}.
This difference implies that 
taking into account only the screening effect due to
forward scattering process is not sufficient for the correct
treatment of the voltage drop.
One needs to consider the renomalization of the external potential
due to other scattering processes.
The theoretical treatment of the voltage drop 
was also discussed by Egger and Grabert 
from another point of view based upon Landauer-type
approach\cite{eg}.
More recently, Kawabata pointed out that corrections to the
conductance due to short-ranged electron-electron interaction 
may be absorbed into the renormalization of a chemical 
potential\cite{kawa}.
He demonstrated this in the case of backward scattering.
By calculating corrections up to the first order in electron-electron
interaction,
he obtained the unrenormalized value of the conductance, $2e^2/h$.
In this paper, we investigate the renormalization of the chemical
potential due to Umklapp scattering 
in the case away from half-filling, 
which has not been
considered in the previous studies.
Our main purpose is to show that we have 
the unrenormalized value of the conductance, $2e^2/h$,
by taking into account both the screening of external
potential due to forward scattering and the renormalization of
the chemical potential due to Umklapp scattering\cite{imp}. 
For this purpose, we need to calculate 
the second order processes of Umklapp scattering 
which give singular contributions.
Thus, in contrast to the case of backward scattering,
we cannot adopt a simple perturbation approach.
We exploit bosonization and renormalization group method.

The organization of this paper is as follows.
In Sec. II, we define our model and 
summarize the results of our previous paper\cite{fk},
in which only the screening of external potential due to forward
scattering is considered.
In Sec. III, we develop perturbative renormalization group
argument for the renormalization of the chemical potential
due to Umklapp scattering. 
Combining the results in Secs. II and III, we show that
the conductance is not renormalized even in the presence of Umklapp
scattering in Sec. IV.
In Sec. V, we consider the generalization of our argument to fermion
systems with SU(N) internal symmetry. 
This generalization is worthwhile to be investigated because of
the following reason. 
It has been recently claimed by Zotos et al. that
all 1D quantum {\it integrable} systems may have
the properties of ideal conductors with
zero-resistivity\cite{zotos} even at finite temperatures.
This assertion has been directly confirmed 
by the calculation based upon the 
Bethe ansatz solutions\cite{fk2,naro}.
In this sense, integrable systems may exhibit rather special 
properties for transport coefficients in contrast
to general non-integrable cases.
Since our argument for the electron model (SU(2) case) is 
based on the bosonized effective hamiltonian, i.e.
the 1D sine-Gordon model, which is integrable,
we need to show that the unrenormalized conductance is not
the consequence of the integrability, but 
a universal property inherent in 1D electron systems
(even with Umklapp interaction).  
To confirm this point, we investigate a non-integrable SU(N)
fermion model with Umklapp scattering. 
In the last section, some discussions about the applicability
of our argument are given.

\section{Screening of the external potential due to
forward scattering}

We consider an  electron system with forward and
Umklapp interactions, 
and start with the following Hamiltonian 
after linearizing the dispersion
around the Fermi points,
\begin{eqnarray}
H&=&
{\rm i}\hbar v_F \int {\rm d}x \sum_{\sigma} :\psi_{\sigma L}^{\dagger}(x)
\partial_x\psi_{\sigma L}(x)-
\psi_{\sigma R}^{\dagger}(x)\partial_x\psi_{\sigma R}(x): \nonumber \\
&+&g\int \frac{{\rm d}x}{2\pi} :\rho(x)\rho(x): \nonumber \\
&+&U\int \frac{{\rm d}x}{2\pi} :{\rm e}^{{\rm i}(4k_{\rm F}-2\pi)}
\psi^{\dagger}_{\uparrow L}(x)\psi_{\uparrow R}(x)
\psi^{\dagger}_{\downarrow L}(x)\psi_{\downarrow R}(x)+{\rm h.c.}:,
\label{hamil1}
\end{eqnarray}
where 
$::$ represents the normal ordering,
$\psi_{\sigma L(R)}$ is the annihilation
operator for left(right)-moving electrons
with spin $\sigma$,
$g$ ($U$) is the coupling for
forward (Umklapp) scattering, 
and $\rho(x)=\rho_L(x)+\rho_R(x)$ with
$\rho_{L,(R)}(x)=\sum_{\sigma}\psi^{\dagger}_{\sigma L(R)}(x)
\psi_{\sigma L(R)}(x)/\sqrt{2}$. 
According to Kawabata,\cite{kawa2,kawa}
the renormalization of the external potential, $\Phi_0(q,\omega)$, 
occurs due to electron-electron interaction.
In the present case, we have two kind of
interactions, both of which may be expected to contribute
to the potential renormalization.
The forward scattering screens the external potential
and changes the measured voltage. 
The Umklapp scattering gives rise to 
the non-uniform chemical potential
shift in the presence of electric field gradient which contributes
to the renormalization of external fields.
The former effect was considered in our previous paper\cite{fk},
which is briefly summarized in this section. 

The screening effect 
can be incorporated by calculating the diagrams for
current-current 
correlation function which are irreducible with
respect to the forward scattering, $g$.
Using bosonization method, we have
the effective Hamiltonian for
the charge degrees of freedom of eq.(\ref{hamil1}),  
\begin{eqnarray}
H&=&H_0+H_u, \\
H_0&=&\int {\rm d}x\biggl
[\frac{v_{\rho}}{2 K_{\rho}}
(\partial_x \phi_{\rho}(x))^2+\frac{v_{\rho}
K_{\rho}}{2}(\Pi_{\rho}(x))^2\biggr] \\
H_u&=&\frac{U}{\alpha^2}
\int {\rm d}x \cos(\sqrt{8\pi}\phi_{\rho}(x)+\delta x), \label{umkl}
\label{bosonham}
\end{eqnarray}
where $\alpha$ is the high-energy cut-off parameter.
Here $\phi_{\rho}$ is a boson 
phase field for the charge degrees of freedom,
$\Pi_{\rho}$ is its canonical conjugate field,
and $\delta\equiv 4k_{\rm F}-2\pi$ with
$k_{\rm F}$ being the Fermi point.
In the following, we will consider only the case 
away from half-filling, $\delta\neq 0$. 

Since the Umklapp scattering term 
becomes irrelevant at the TL liquid fixed point,
the leading correction to the conductance due to
the Umklapp term can be estimated using perturbative calculations.
The conductance is given by
\begin{eqnarray}
G=\lim_{\omega\rightarrow 0}
\frac{e^2\bar{\omega}^2}{L^2\omega}
\int^{L/2}_{-L/2} {\rm d}x\int^{L/2}_{-L/2} {\rm d}x' 
\int^{\beta}_{0}
d\tau\langle T\phi_{\rho}(x,\tau)\phi_{\rho}(x', 0)\rangle
e^{-i\bar{\omega}\tau}\vert_{\bar{\omega}=i\omega-\varepsilon}.
\label{condge}
\end{eqnarray}
In order to take into account the renormalization of
external potential, we should calculate the irreducible diagram
with respect to the forward scattering, $g$,
which is related to the charge susceptibility
$\chi(q,\omega)$,
\begin{equation}
q^2\langle\phi_{\rho}(q,\omega)\phi_{\rho}(-q,\omega)\rangle^R_{\rm irr}
=\frac{\chi(q,\omega)}{1-\frac{g}{2}\chi(q,\omega)},
\label{chid}
\end{equation}
for $q\sim 0$. Here, $\langle\cdot\cdot\cdot\rangle^R$
is the retarded Green's function.
We now expand $\chi(q,\omega)$ in terms of the strength of the Umklapp
scattering $U$:
$\chi(q,\omega)=\tilde{\chi}(q,\omega)+\delta\chi(q,\omega)$,
where $\tilde{\chi}(q,\omega)$ 
includes only the effect of the forward scattering
whereas $\delta\chi(q,\omega)$ is the correction due to the Umklapp term.
Then we have
\begin{eqnarray}
q^2\langle\phi_{\rho}(q,\omega)\phi_{\rho}(-q,\omega)\rangle^R_{\rm irr}
&=&\frac{\tilde{\chi}(q,\omega)}
{1-\frac{g}{2}\tilde{\chi}(q,\omega)}
+\frac{\delta\chi(q,\omega)}
{(1-\frac{g}{2}\tilde{\chi}(q,\omega))^2} \nonumber \\
&=&\chi_0(q,\omega)+
\Bigl(\frac{\chi_0(q,\omega)}{\tilde{\chi}(q,\omega)}\Bigr)^2
\delta\chi(q,\omega),
\label{chicor}
\end{eqnarray}
where $\chi_0(q,\omega)$ is the charge susceptibility for non-interacting
electron systems.
Evaluating $\delta\chi(q,\omega)$ up to the second order in $U$,
we have the renormalized conductance\cite{fk},
\begin{equation}
G=\frac{2e^2}{h}(1-bU^2)
\label{condum1}
\end{equation}
with $b=K_{\rho}(e^{(4-4K_{\rho})l_c}-1)/(8-8K_{\rho})$.
Here $l_c$ is determined by the condition that 
$|4k_{\rm F}-2\pi|\sim 1/\alpha e^{l_c}$ where $\alpha$ 
is a high-energy cutoff\cite{gia}.
Note again that the formula (\ref{condum1}) is obtained
by incorporating the screening effect due to 
forward scattering.
This result should be modified, if one 
further takes into account properly
the renormalization of the chemical potential due to 
Umklapp scattering, which will be discussed 
in the following sections.

\section{Renormalization of the chemical potential
due to Umklapp scattering}

In this section, we discuss the renormalization of the non-uniform
chemical potential due to the Umklapp scattering which results in
the renormalization of the external electric fields.
We use the perturbative renormalization group method.
The coupling of charge currents to the external potential
$\Phi(x,t)$  is given by
\begin{equation}
H_{\rm ext}=-\sqrt{\frac{2}{\pi}}
\int {\rm d}x \Phi(x,t)\partial_x\phi_{\rho}(x).
\end{equation}
Here, for simplicity and making our argument clear, we
omit for a while the screening effect of external potential due to
the forward scattering, and concentrate on the renormalization
of the chemical potential due to the Umklapp scattering.
Both effects are considered in the next section to 
obtain the final formula for the conductance.

Up to the first order in $\Phi(x,t)$ and the second order in $U$, 
we obtain the renormalization group equation for the external
potential,
\begin{eqnarray}
\frac{{\rm d} \Phi(x,t)}{{\rm d} l}
&=&\Phi(x,t)-2\pi\frac{U^2 K_{\rho}}{v_{\rho}^2}
J_0(\delta\alpha)\Phi(x,t) \nonumber \\
&&+2\pi\frac{U^2K_{\rho}}{v_{\rho}^2}
\frac{J_1(\delta\alpha)}{\delta\alpha}\Phi(x,t),
\label{phirg}
\end{eqnarray}
where $J_n(x)$ is the Bessel function.

Note that the renormalization equation 
for the uniform chemical potential is given by
\begin{equation}
\frac{{\rm d} \delta}{{\rm d}
l}=\delta+2\pi\frac{U^2K_{\rho}}{v_{\rho}^2}
\frac{J_1(\delta\alpha)}{\alpha},
\label{delrg}
\end{equation}
where the quantity $\delta$ has been introduced in 
eq.(\ref{bosonham}). 
Thus eq.(\ref{phirg}) is rewritten as,
\begin{equation}
\frac{{\rm d} \Phi(x,t)}{{\rm d} l}=\frac{1}{\delta}
\frac{{\rm d} \delta}{{\rm d} l}\Phi(x,t)
-2\pi\frac{U^2 K_{\rho}}{v_{\rho}^2}
J_0(\delta\alpha)\Phi(x,t).
\end{equation}
We impose the condition that ${\rm d}\delta/{\rm d} l=0$
in order to conserve the electron density.
Then we have
\begin{equation}
\frac{{\rm d} \Phi(x,t)}{{\rm d} l}=-2\pi\frac{U^2
K_{\rho}}{v_{\rho}^2}
J_0(\delta\alpha)\Phi(x,t).
\label{phirg2}
\end{equation}

The current induced by the bare external potential $\Phi_0(x,t)$
at zero temperature is $I=(2e^2/h)K_{\rho}\Delta\Phi_0$ with 
$\Delta\Phi_0=\Phi_0(+\infty, t)-\Phi_0(-\infty, t)$.
Thus the conductance is given by
\begin{equation}
G=\frac{2e^2}{h}\frac{K_{\rho}\Delta\Phi_0}{\Delta\Phi}.
\end{equation}
Using eq.(\ref{phirg2}) and the renormalization equation of
$K_{\rho}$\cite{gia},
\begin{equation}
\frac{{\rm d} K_{\rho}}{{\rm d}
l}=-2\pi\frac{U^2K_{\rho}^2}{v_{\rho}^2}
J_0(\delta\alpha),
\end{equation}
we have,
\begin{equation}
\frac{{\rm d} G}{{\rm d} l}=0.
\label{nonrenorm}
\end{equation}
Thus the conductance is not renormalized by the Umklapp scattering
for any scaling length at zero temperature.
This remarkable property provides the basis for the following argument
about the universal conductance in the presence of the Umklapp
scattering.  To avoid confusions, we wish to mention 
again that the formula (\ref{nonrenorm}) has been obtained 
by omitting the potential-renormalization effect due to
forward scattering for simplicity.

\section{Universal conductance in the presence of Umklapp scattering}

Now we are ready to show that the conductance is not renormalized 
even in the presence of Umklapp scattering if one incorporates 
both effects of the screening of external potential
due to forward scattering and the renormalization of the chemical
potential due to Umklapp scattering.

The averaged value of the current in the static limit is given by
\begin{eqnarray}
\langle J(x)\rangle &=&\lim_{\omega\rightarrow 0}
\int dq\frac{1}{q}\chi(q,\omega)\Phi_0e^{iqx} \nonumber \\
&=&\lim_{\omega\rightarrow 0}\int dq\frac{1}{q}\tilde{\chi}(q,\omega) 
\Phi_0(q,\omega)e^{iqx}+\frac{2e^2}{h}
\delta K_{\rho}\Delta \Phi_0, \nonumber \\
&\equiv& J_1(x)+J_2(x),
\label{cur4}
\end{eqnarray}
where we have expanded the charge susceptibility $\chi(q,\omega)$
in terms of $U$, and separated the current into
two parts: $J_1(x)$, which includes only the effect of 
the forward scattering,
and $J_2(x)=(2e^2/h)\delta K_{\rho}\Delta\Phi_0$, 
the correction due to the Umklapp scattering. 
This expansion is justified for small $U$ since the Umklapp interaction is
an irrelevant operator. 
We also expand the external field in terms of $U$,
\begin{equation}
\Phi'(x,t)=\Phi(x,t)+\delta\Phi(x,t),
\end{equation}
where $\Phi(x,t)$ includes only the screening effect due to the forward
scattering, and $\delta\Phi(x,t)$ is the correction
due to the Umklapp scattering discussed in Sec. III.

Then the contribution to the conductance
from the first term of the last line in eq.(\ref{cur4})
is given by up to the lowest order in $U$,
\begin{eqnarray}
G_1\equiv\frac{\int dx J_1(x)}{\Delta\Phi'}
&=&\frac{2e^2}{h}-\frac{2e^2}{h}\tilde{K}_{\rho}
\frac{\Delta\Phi_0\Delta(\delta\Phi)}{(\Delta\Phi)^2}+O(U^3).
\label{gg1}
\end{eqnarray}
Here $\tilde{K}_{\rho}$ includes only the contribution
from forward scattering, and 
$\Delta(\delta\Phi)=\delta\Phi(+\infty)-\delta\Phi(-\infty)$.
In order to obtain this expression, we have used the relation
$\chi_0(q,\omega)\Phi=\tilde{\chi}(q,\omega)\Phi_0$\cite{kawa2,fk}.
The contribution to the conductance from 
$J_2(x)$ up to the same order in $U$ is given by,
\begin{equation}
G_2\equiv\frac{\int dx J_2(x)}{\Delta\Phi'}\simeq
\frac{2e^2}{h}\frac{\delta K_{\rho}\Delta\Phi_0}{\Delta\Phi}.
\label{gg2}
\end{equation}
The corrections due to Umklapp scattering, $\Delta(\delta\Phi)$
and $\delta K_{\rho}$, are evaluated by using 
the renormalization group equation obtained in Sec. III,
\begin{equation}
\Delta(\delta\Phi)=\int dl\frac{d\Delta\Phi}{dl}=
-2\pi\int dl \frac{K_{\rho}U^2}{v_{\rho}^2}J_0(\delta\alpha)
\Delta\Phi,
\label{dp}
\end{equation}
\begin{equation}
\delta K_{\rho}=-2\pi\int
dl\frac{K_{\rho}^2U^2}{v_{\rho}^2}J_0(\delta\alpha).
\label{kr}
\end{equation}
If we consider the lowest order corrections in $U$,
$K_{\rho}$ in the right-hand side of eqs.(\ref{dp})
and (\ref{kr}) is replaced by $\tilde{K}_{\rho}$, and
$\Delta\Phi$ in the right-hand side of eq.(\ref{dp}) 
does not depend on the scaling parameter
$l$. 
Then from eqs.(\ref{gg1}) and (\ref{gg2}), we have
the conductance,
\begin{equation}
G=G_1+G_2=\frac{2e^2}{h}.
\end{equation}
Therefore we come to the conclusion that
the conductance is not renormalized even in the presence of Umklapp
scattering if one takes into account
not only the screening of the external potential but also
the renormalization of the local chemical potential. 
This conclusion, which may improve our previous results\cite{fk},
 is in accordance with the recent studies
on the effect of non-interacting leads attached to quantum wires
in the presence of Umklapp scattering\cite{mori,mas}.

\section{Generalization to the SU(N) fermion model}

As mentioned in the introduction, the model we considered in the
previous sections, the sine-Gordon model, is an integrable
system, in which it is known that no true current decay
occurs\cite{zotos}. 
Thus one may suspect that the unrenormalized conductance
obtained in the previous sections might be the consequence of
the integrability of the model.
In this section, in order to examine this point, 
we consider the non-integrable 1D interacting electron systems
of which the spin degrees of freedom is generalized to the
SU(N) symmetry.
In this model, the bosonized form of the Umklapp interaction 
which breaks the integrability
is given by\cite{aff},
\begin{equation}
H_{umklapp}=\frac{U}{\alpha^2}\int dx (g^{\alpha}_{\beta}g^{\beta}_{\alpha}-
g^{\alpha}_{\alpha}g^{\beta}_{\beta})\exp(i\sqrt{16\pi/N}\phi_{\rho}
+\delta x)
+h.c.,
\end{equation}
where $g^{\alpha}_{\beta}$ is a matrix 
of SU(N) Lie group.
As seen from the above interaction, the charge degrees of freedom
is coupled with SU(N) internal degrees of freedom.  
In the case away from half-filling $\delta\neq 0$,
this term is irrelevant, and can be treated perturbatively.
Following the method in Sec. III,
we obtain the scaling equations for external potential and
the Luttinger parameter $K_{\rho}$,
\begin{equation}
\frac{{\rm d} \Phi(x,t)}{{\rm d} l}=-4\pi\frac{U^2
K_{\rho}}{N v_{\rho}^2}
J_0(\delta\alpha)\Phi(x,t),
\label{sunphi}
\end{equation}
\begin{equation}
\frac{{\rm d} K_{\rho}}{{\rm d}
l}=-4\pi\frac{U^2K_{\rho}^2}{N v_{\rho}^2}
J_0(\delta\alpha).
\end{equation}
Using these equations, we can repeat the same argument
as done for the SU(2) case, and obtain the unrenormalized value of
the conductance, $2e^2/h$.
Thus, our result is not restricted to integrable systems.
Although our argument is based upon a specific model,
we believe that the result for the unrenormalized conductance
in the presence of Umklapp scattering 
is a universal property of 1D metallic systems.

\section{Discussions}

In this paper, we have shown that 
the conductance takes universal value, $2e^2/h$,
in the presence of Umklapp scattering, by 
properly taking into account not only 
the screening of external potential due to forward scattering
of electron-electron interaction but also
the renormalization of chemical potential due to Umklapp
scattering.  The conclusion is in accordance with the 
theoretical result obtained for 
the system with  non-interacting leads being attached
to a quantum wire, although
the mechanism to obtain the universal value is 
different between two approaches.
Here we discuss about the condition in which
our argument is applicable. 
The renormalization of the local chemical potential stems from
the local charge density fluctuation induced by
the external potential.
In the derivation of eq.(\ref{phirg}), we assumed that
the induced charge density, 
$\chi_c\Phi(x,t)=(2K_{\rho}/\pi v_{\rho})\Phi(x,t)$
is much smaller than the total charge density $n$.
This is nothing but the condition required for the applicability of
linear response theory.
As the electron density approaches half-filling $n\rightarrow 1$,
the charge susceptibility $\chi_c$ diverges\cite{usuki} like 
$\sim 1/(1-n)$.
Thus in the vicinity of the Mott transition,
the value of $\Phi(x,t)$ for which the unrenormalized conductance is
observed is quite small, $\Phi(x,t)\ll n/\chi_c$.
We cannot apply Landauer's formula 
unless this condition is satisfied.
Thus although the conductance is  not renormalized
for any electron densities away from half-filling, 
the range of the applied external potential 
for which the unrenormalized conductance is observable becomes smaller, 
as the electron density approaches the half-filling.
A sufficiently large external potential
may excite electrons to the upper-Hubbard band.
In this case, non-linear effects become very important.
It may be an interesting issue to investigate
such non-linear effects which characterize
the precursor of the Mott transition. 

\acknowledgements{}

One of the authors (S. F.) thanks F. Essler, D. Shelton, and
A. M. Tsvelik for illuminating conversation and their hospitality
at University of Oxford.
This work was partly supported by a Grant-in-Aid from the Ministry
of Education, Science, Sports and Culture, Japan.



\begin{references}

\bibitem{hal} F. D. M. Haldane, J. Phys. C {\bf 14}, 2585 (1981);
Phys. Rev. Lett. {\bf 47}, 1840 (1981).

\bibitem{ap} W. Apel and T. M. Rice, Phys. Rev. B{\bf 26}, 7063 (1982).

\bibitem{kane} C. L. Kane and M. P. A. Fisher, Phys. Rev. Lett.
{\bf 68}, 1220 (1992).

\bibitem{taru} S. Tarucha, T. Honda, and T. Saku, Solid State Commun.
{\bf 94}, 413 (1995).

\bibitem{stone} D. L. Maslov and M. Stone, Phys. Rev. B{\bf 52}, 5539 (1995).

\bibitem{pome} V. V. Ponomarenko, Phys. Rev. B{\bf 52}, 8666 (1995).  

\bibitem{schu} H. Safi and H. J. Schulz, Phys. Rev. B{\bf 52}, 
17040 (1995).

\bibitem{kawa2} A. Kawabata, J. Phys. Soc. Jpn. {\bf 65}, 30 (1996).

\bibitem{fink} Y. Oreg and A. M. Finkel'stein, Phys. Rev. B{\bf 54},
14265 (1996)

\bibitem{shimi} A. Shimizu, J. Phys. Soc. Jpn. {\bf 65}, 1162 (1996).

\bibitem{maslov} D. L. Maslov, Phys. Rev. B{\bf 52}, 14368 (1995).

\bibitem{furu} A. Furusaki and N. Nagaosa, Phys. Rev. B{\bf 54},
5239 (1996).

\bibitem{fk} S. Fujimoto and N. Kawakami, J. Phys. Soc. Jpn. {\bf 65},
3700 (1996).

\bibitem{naga} V. V. Ponomarenko and N. Nagaosa, Phys. Rev. Lett. 
{\bf 77}, 1714 (1997); preprint, cond-mat/9711167.

\bibitem{gra} A. Gramada and M. E. Raikh, Phys. Rev. B{\bf 55}, 1661 (1997).

\bibitem{toku} A. A. Odintsov, Y. Tokura, and S. Tarucha,
Phys. Rev. B{\bf 56}, 12729 (1997).

\bibitem{mori} M. Mori, M. Ogata, and H. Fukuyama, J. Phys. Soc. Jpn.
{\bf 66}, 3363 (1997).

\bibitem{mas}  O. A. Starykh and D. L. Maslov, Phys. Rev. Lett. 
{\bf 80}, 1694 (1998).

\bibitem{safi} I. Safi, Phys. Rev. B{\bf 55}, 7331 (1997).

\bibitem{eg} R. Egger and H. Grabert, Phys. Rev. Lett. {\bf 77},
538 (1996).

\bibitem{kawa} A. Kawabata, preprint, cond-mat/9701171.

\bibitem{imp} It is easily seen that 
the renormalization of the non-uniform chemical potential 
due to impurity scattering does not exist.
Thus a periodic impurity potential may give the non-universal value of
the conductance in our framework.

\bibitem{zotos} X. Zotos and P. Prelovsek, Phys. Rev. B{\bf 53}, 983
(1996); X. Zotos, F. Naef, and P. Prelovsek, Phys. Rev. B{\bf 55},
11029 (1997).

\bibitem{fk2} S. Fujimoto and N. Kawakami, J. Phys. A{\bf 31}, 465 (1998).

\bibitem{naro} B. N. Narozhny and N. Andrei, preprint, 
cond-mat/9711100.

\bibitem{gia} T. Giamarchi, Phys. Rev. B{\bf 44}, 2905 (1991).

\bibitem{aff} I. Affleck, Nucl. Phys. B{\bf 305}, 582 (1988).

\bibitem{usuki} T. Usuki, N. Kawakami, and A. Okiji,
J. Phys. Soc. Jpn. {\bf 59}, 1357 (1990).

\end{references}
\end{document}